\documentstyle[preprint,eqsecnum,aps,epsf]{revtex} 
   
\begin{document}    
\preprint{\large{\sl Submitted to Phys.\,Rev.\,B}} 
\title{\Large {\bf Volume Effect in the Landau Theory of Martensitic    
	Phase Transitions in Cubic Crystals}}     
   
\author {M.A.Fradkin\cite{Carleton}}   
   
 \address{Institute of Crystallography, Russian Academy of Sciences \\   
	59 Leninsky prospect, Moscow, 117333, Russia}
   
\date{June 5, 1994} 
   
\draft   

\hyphenation{ma-r-te-n-si-tic}   
\hyphenation{Pre-v-ra-sche-nij}   
\hyphenation{So-u-n-ders}

\maketitle    

\thispagestyle{empty}   
   
\begin{abstract}    
	An effect of the volume change upon proper ferroelastic 	   
	(martensitic) phase transitions in cubic crystals is considered.    
	Corresponding terms in the Ginzburg-Landau expansion of the Gibbs    
	free energy are analyzed for the first- as well as second-order phase 
	transitions from cubic to tetragonal lattice under the action of 
	uniaxial and hydrostatic pressure. The 
	pressure effect on the critical	temperature as well as on the phase 
	transition anomalies of isothermal compressibility and linear thermal 
	expansion coefficient are studied and recent experimental data on 
	thermal expansion anomalies in V$_3$Si, In-Tl and Ni-Al are discussed. 
	The non-linearity of thermal expansion 
	leads to the special relation between the shear strain and volume 
	change as a result of the elastic energy minimization. This phenomenon
	can provide the transformation from FCC lattice to BCC one, 
	observed in the iron alloys.   
\end{abstract}    
    
\bigskip

   
\section{Introduction}   
	There is a growing attention in recent years to the physics    
	of martensitic phase transformations in metal    
	alloys\cite{Tanner,Shapiro}, though these phenomena were known by    
	materials scientists for many years as diffusionless    
	transformations characterized by specific transformation    
	kinetics\cite{Christian} not described by the classical    
	nucleation theory\cite{Lubov}. Two different kinds of martensitic    
	transformations are known, i.e., an athermal and isothermal ones.   
	In the athermal case the transformation begins at some start    
	temperature $M_s$, but the parent phase still exists until the 
	temperature goes down to $M_f$ - a martensite    
	finish point. In the isothermal case the transformation proceeds    
	in time at a constant temperature and, generally speaking, could    
	be completed in some finite time, which might be very long and    
	depends in turn on temperature. There is    
	a particular sub-class of a so-called "thermo-elastic" martensite    
	within this class of transformation, which is characterized by the    
	reversibility of a structure change - the alloy regains its   
	high-temperature structure upon heating from a martensitic   
 	low-temperature state through the transition point. This means that    
	the lattice of product phase is coherent with respect to the    
	parent one. This phenomenon gives the ground for the "shape-memory    
	effect"\cite{AMSHandb} and is, thus, of a great practical    
	importance.    
   
	Due to a spontaneous strain release during the transformation some   
	elastic distortions appear in the matrix of the parent phase    
	surrounding the martensite nuclei. The minimization of elastic   
        long-range energy determines the shape of new phase precipitates   
        and provides an equilibrium state that has complicated  
	heterogeneous structure which involves multiple-twin bands for  
	deeper relaxation of the elastic strain\cite{Royt}. Thus, a real    
	development of transformation takes place in complicated    
	conditions of {\em apriory} unknown external pressure from the    
	lattice of the parent phase upon the regions under    
	transformations. Thus, the heterogeneity of the system makes it   
	difficult to analyze the thermodynamic of phase transition, and    
	idealized single-crystal systems should be considered in order to    
	study an equilibrium structure of the martensite phase as well as    
	an equilibrium development of the transformation.     
   
	The reversibility of the transformation means that a phase    
	transition between equilibrium phases takes place and the structure 
	coherence as well as the symmetry breaking   
 	at a critical temperature implies that the transition can be    
	analyzed within the frame of the Landau theory. In this approach    
	the difference in free energy between the parent and product phase    
	is considered as a function of some order parameter which is equal    
	to zero in a high-temperature symmetrical phase and becomes    
	non-vanishing below the critical temperature\cite{LandLif}. The    
	fundamental difference of the structure of a martensitic phase    
	with respect to the parent lattice is well known to be spontaneous    
	strain\cite{Royt}, so this phase transition belongs to the class    
	of proper ferroelastics\cite{Salje}. It means that the structure    
	change takes place through the elastic instability of the    
	crystalline lattice of a parent high-temperature phase with    
	respect to a spontaneous homogeneous strain of a special    
	kind\cite{Warlimont}. In other words, some combination of the    
	elastic modulii vanishes at the critical temperature.    
   
	It gives rise to the drastic lowering of the frequency of a    
	corresponding mode of the acoustic vibration leading to a central    
	peak of the inelastic neutron scattering\cite{SMSHAP} and this is    
	the reason why this sort of phase transition is often referring to    
	as a "soft-mode" one. Though, the modes never become completely   
	soften\cite{GodKru-89} and a finite (albeit -- small) jump of the    
	order parameter is usually observed, the martensitic phase    
	transition still may be treated as a weakly-discontinuous and    
	considered in the frame of the Landau theory of a continuous phase    
	transition. A corresponding Ginzburg-Landau expansion of the free    
	energy in series of the symmetrized strain components has been    
	developed\cite{Bocc,Liak,Falk} and the heterogeneous    
	fluctuations\cite{Cao}, nucleation of the martensite phase around    
	the defects\cite{Reid} and some other phenomena\cite{Pouget} were    
	studied in such a formalism.    
   
	In the present paper the role of hydrostatic strain in the    
	martensitic phase transitions is considered for the case of a    
	cubic lattice of a high-temperature phase. A cubic point symmetry    
	leads to a coupling between the shear strain and volume change    
	in the elastic energy expansion near the critical temperature, and    
	the volume change can be considered as an additional order    
	parameter not related with the symmetry breaking. The coupling is    
	shown to lead to the transition anomaly in the thermal    
	expansion coefficient as well as in the isothermal 	   
	compressibility. For the case of the first-order transition some    
	effect is found to occur in the low-temperature phase even at    
	the temperature region outside fluctuation-induced singularities.    
	   
	The effect of the uniaxial pressure, that preserves the symmetry of   
	the low-temperature tetragonal phase is studied as well. It is an    
	external field conjugated to the order-parameter which is    
	known\cite{LandLif} to suppress the continuous (second-order)    
	phase transition for arbitrary small field value, however, the    
	weakly-discontinuous transition is preserved under the field lower    
	than the critical one. The dependence of the transition    
	temperature on the external uniaxial as well as hydrostatic    
	pressure is derived and the critical pressure is found.    
   
	The non-linear analysis of the volume change shows, that the    
	elastic energy minimization can provide the FCC structure of  
	low-temperature phase for the BCC parent lattice through the  
	special relationship between the shear strain and volume change. 
	Hence, non-linear elastic effects might be responsible for the    
	challenging FCC -- BCC martensitic transformation in pure iron and    
	some ferrous alloys.   
	   
	The paper is organized as follows. A brief description of a    
	weakly-first-order transition within the Landau theory is a    
	content of the Section \ref{GLT}. The Landau theory of a proper    
	ferroelastic (martensitic) transformation is analyzed in Section    
	\ref{PFE}. The volume change due to such a phase transition    
	is considered in Section \ref{VLC}. Non-linear thermal expansion    
	and its possible contribution to the martensitic phase transition    
	is studied in Section \ref{BCCFCC}.

\section{The Landau theory of the first-order phase transition}   
	\label{GLT}   
   
\subsection{Continuous phase transition}   
	\label{second-order}   
   
	Let us recall briefly main ideas of the Landau theory of    
	continuous phase transitions\cite{LandLif}. As Landau supposed, if    
	the symmetry group of low-symmetry phase $G_1$ is a subgroup of    
	the symmetry group $G_0$ of the high-symmetry one, than there is    
	some variable $\eta$, called as an "order parameter" which is    
	invariant under all the transformations from the $G_1$ whereas    
	some transformations from $G_0$ change it. The thermodynamic    
	potential -- the Gibbs free energy could be expanded as a power    
	series in $\eta$ near the critical temperature. As the    
	thermodynamic potential should not change under the symmetry    
	transformations which do not change the structure, the $\eta$ in  
	the high-temperature phase should vanish.    
   
	General expression for the (Ginzburg-Landau) expansion of the    
	difference in Gibbs free energy between the phases has the    
	following form\cite{LandLif}   
\begin{equation}   
	\Delta {\cal G} = {\alpha \over 2}(T - T_c)\eta^2 +    
	{C \over 4}\eta^4  
	\label{GLsecond}  
\end{equation}   
	where $T_c$ is a critical temperature and the coefficients    
	$\alpha$ and $C$ should be positive. The equilibrium value of    
	$\eta$ is determined by the minimization of $\Delta {\cal G}$ with    
	respect to $\eta$:   
\begin{equation}   
	\frac{\partial \Delta {\cal G}}{\partial \eta} = 0    
	\hspace{12pt} {\rm and} \hspace{12pt} \ \ \   
	\frac{\partial^2 \Delta {\cal G}}{\partial^2 \eta} > 0.   
	\label{minimum}   
\end{equation}   
	The solutions are the high-symmetry phase with $\eta = 0$, stable    
	for $T > T_c$ and low-temperature phase with $\eta^2 = \alpha    
	(T_c-T)/C$ that is stable for $T < T_c$. The dependence of $\eta$    
	on $T$ is continuous in the critical point $T_c$, hence this model    
	describes the second-order phase transition. The Gibbs free energy    
	as well as entropy changes continuously trough the transition  
	temperature $T_c$   
\begin{equation}   
	\Delta {\cal G} = - {\frac{\alpha^2}{4 C}}(T - T_c)^2    
\end{equation}   
\begin{equation}   
	\Delta {\cal S} = - \frac{\partial \Delta {\cal G}}{\partial T} =    
	 		  	{\frac{\alpha^2}{2 C}}(T - T_c) \ ,   
\end{equation}   
	but the derivative manifests discontinuity   
\begin{equation}   
	\Delta {\cal C}_P = T \frac{\partial \Delta {\cal S}}{\partial T}    
		= - T \frac{\partial^2 \Delta {\cal G}}{\partial T^2} =    
	 	\frac{\alpha^2 T_c}{2 C}.   
	\label{specheat}  
\end{equation} 
  
	Vanishing of the coefficient of second degree in the Ginzburg-Landau  
	expansion (\ref{GLsecond}) when the temperature approaches $T_c$ leads 
	to the critical fluctuations of the order parameter. Mean square of 
	the homogeneous order parameter fluctuations is given by well-known  
	expression\cite{LandLif}   
\begin{equation}	   
	\langle \eta^2 \rangle \propto {k_B T \over \alpha\,|T - T_c|^{-1}}.   
	\label{fluct}  
\end{equation}  
	Inhomogeneous fluctuations appear to be crucial in many cases,  
	however, for the purposes of present study they can be neglected due 
	to long-range nature of elastic interactions in solids\cite{Cowley}. 
   
\subsection{External field effect}   
	\label{ExtField}   
   
\begin{figure} 
\epsffile{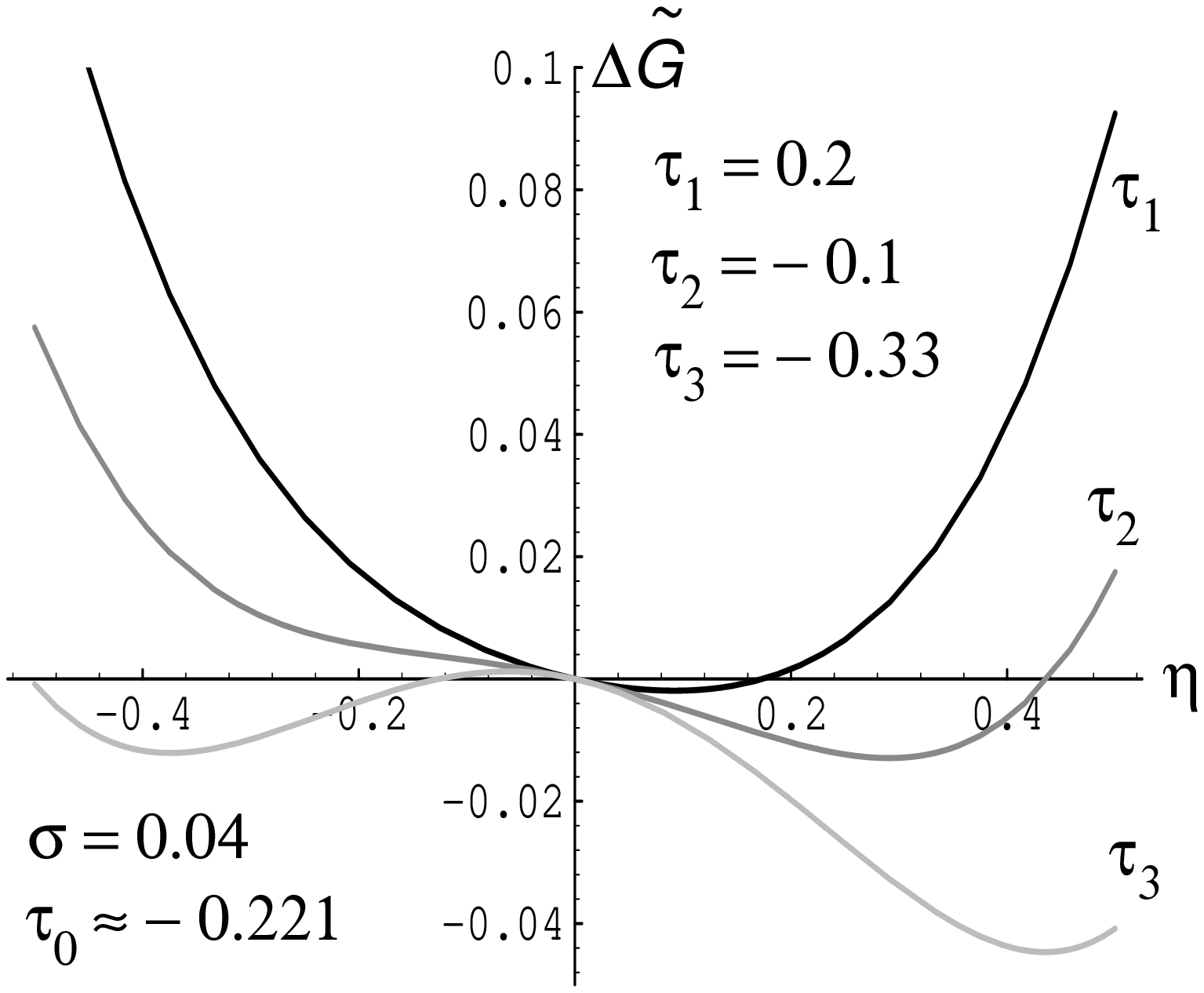} 
\caption{The dependence of the Gibbs energy on the order parameter $\eta$  
        under the applied field for different  
        temperatures $\tau_1 > \tau_2 > \tau_0 > \tau_3$ in the case of  
        the second-order phase transition.}  
\label{IIenergy} 
\end{figure}

\begin{figure} 
\epsffile{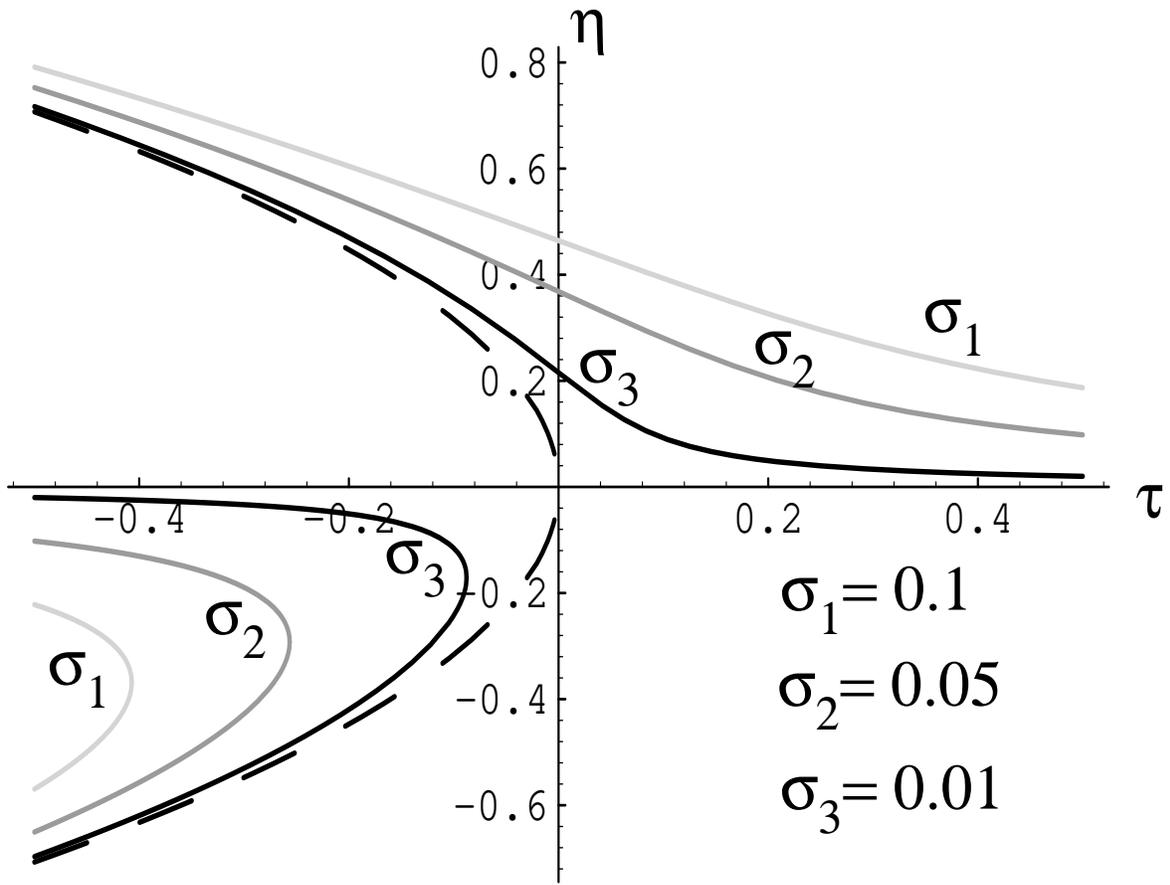} 
\caption{The order parameter dependence on the temperature in various fields 
        for the case of the second-order phase transition. Dashed line  
        corresponds to the absence of external field, $\sigma = 0$.}   
\label{IIordpar} 
\end{figure}

	Let us consider the effect of the external field $E$ conjugated to    
	the physical variable of the order parameter. In what follows it    
	is an external pressure conjugated to the symmetrized spontaneous    
	strain. The Ginzburg-Landau expansions takes the form   
\begin{equation}   
	\Delta {\cal G} = {\alpha \over 2}(T - T_c)\eta^2 +    
		{C \over 4}\eta^4 - E \eta   
	\label{GLF}   
\end{equation}   
	and the minimization of $\Delta {\cal G}$ with respect to $\eta$    
	leads to the cubic equation   
\begin{equation}   
	\frac{\partial \Delta {\cal G}}{\partial \eta}   
	 	= \alpha (T - T_c) \eta + C \eta^3 - E  = 0   
	\label{field}   
\end{equation}   
	with discriminant    
\begin{equation}   
        Q = {\left(\frac{\alpha (T - T_c)}{3 C} \right)}^3 +  
                {\left(\frac{E}{2 C} \right)}^2  
\end{equation}   
	It is seen that for any value of external field $E$ the   
	high-symmetry phase with $\eta = 0$ no longer provides the stable    
	solution of the Eq.(\ref{minimum}). Instead, we get $\eta \neq 0$   
	for any temperature. It is known\cite{Korn}, that the cubic    
	equation has one solution in real numbers for $Q > 0$ and three    
	ones for the case of $Q < 0$. It means that for temperatures below    
	some critical one that now depends on $E$    
\begin{equation}   
	T_0 (E) = T_c - {3 \over {2 \alpha}}    
			{\left({2 C E^2} \right)}^{1 \over 3} 	   
\end{equation}   
        the additional minimum of the Gibbs free energy appears that    
	corresponds to new phase. However, the initial phase described by    
	the high-temperature solution of (\ref{field}) provides the    
	minimum with lower value of the free energy and is, thus, stable.    
	The free energy behavior as well as the order parameter    
	dependencies on the temperature for different values of the    
	external field are shown in the Fig.\ref{IIenergy} and    
	Fig.\ref{IIordpar}.   
   
        It is seen from the Fig.\ref{IIenergy} and might be proven rigorously  
	that different minima of the $\Delta {\cal G} (\eta)$ curve have    
        different energies for any temperature $T < T_0$. Thus, the    
        high-temperature state remains stable throughout all the region of    
        the phase co-existence. Only the condition of $E = 0$ leads to    
        the degeneracy with respect to sign of $\eta$ that implies the   
 	equal energies of different minima. It leads to a phase transition    
	of the first order under the variation of external field at    
	constant temperature $T < T_c$. In other words, the variation of   
	temperature and external field act on the systems described by the 
	expression (\ref{GLF}) in a different way, because the field variation 
	may lead to the phase change but the temperature one may not.  
 
\begin{figure} 
\epsffile{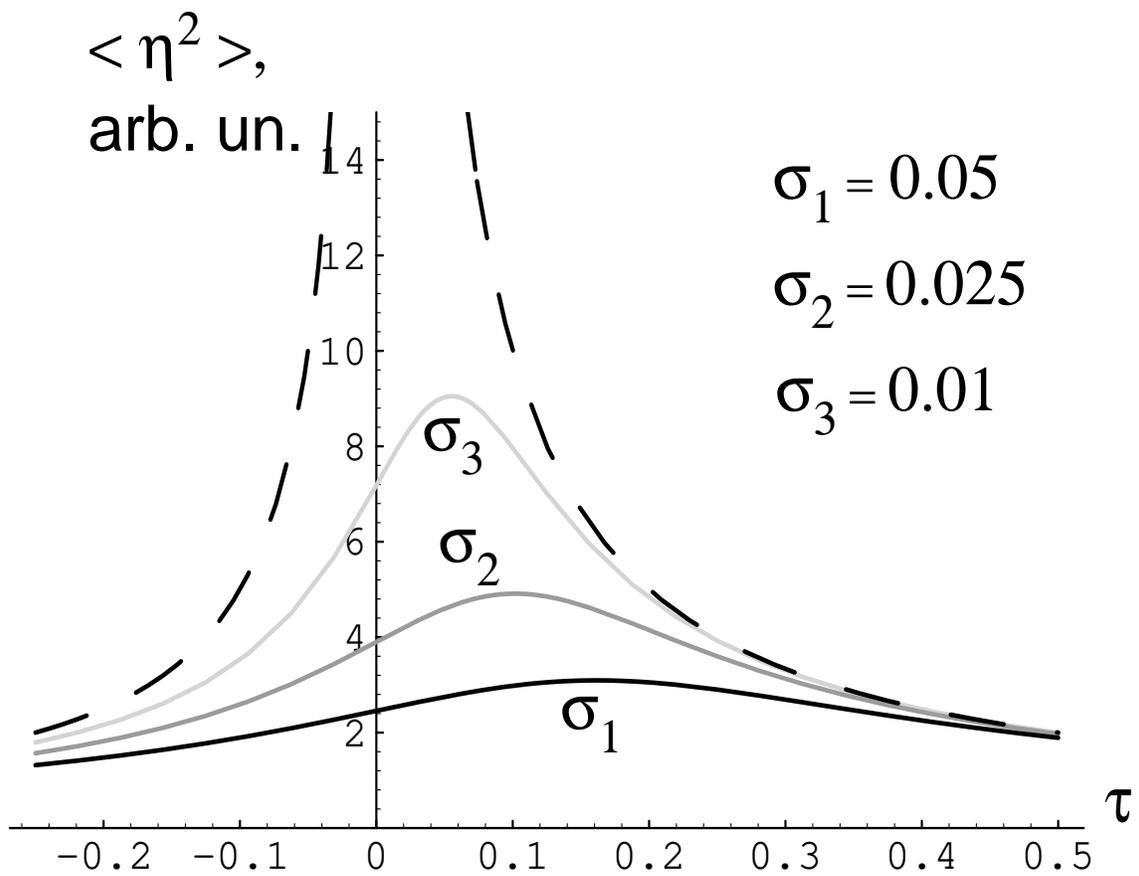} 
\caption{Mean square of the homogeneous order parameter fluctuations around 
	$T_c$ ($\tau = 0$) under different external fields.}  
        \label{fieldfluct}  
\end{figure}

	The external field suppress the transition, however some decrease in 
\begin{displaymath} 
	\frac{\partial^2 \Delta {\cal G}}{\partial \eta^2} (T,E) 
	 	= \alpha (T - T_c) + 3 C \eta^2(T,E) 
\end{displaymath} 
	leads to enhanced fluctuations of the order parameter around $T_c$ 
	with smooth peak instead of divergence (\ref{fluct}) shown in the  
	Fig.(\ref{fieldfluct}). So, some transition anomalies are preserved 
	in sufficiently small external fields $E$.  
   
\subsection{Weakly-discontinuous phase transition in the Landau theory}   
	\label{WFOT}   
 
	The first-order phase transition arises in the Landau theory when    
	the symmetry of the system allows to have non-vanishing third-degree  
	invariant composed by the order-parameter    
	component\cite{LandLif,Toledano}. Corresponding term should be    
	taken into account in the Ginzburg-Landau expansion:   
\begin{equation}   
	\Delta {\cal G} = {\alpha \over 2}(T - T_c)\eta^2 +   
		{B \over 3}\eta^3  + {C \over 4}\eta^4 - E \eta .   
	\label{TOT}   
\end{equation}   
	Choosing the case of $B < 0$, that gives positive $\eta$ in the  
	low-temperature phase, we can write the free energy expansion in the  
	following form   
\begin{equation}   
        \Delta \tilde{\cal G} = {\frac{C^3}{B^4}} \Delta {\cal G} =   
		{\tau \over 2} \ \zeta^2 - {\zeta^3 \over 3}   
		+ {\zeta^4 \over 4} - \sigma \zeta \ ,    
	\label{firstfield}   
\end{equation}   
	with \[ \eta = - {B \over C} \,\zeta \ , \ \   
        \tau = {{\alpha C} \over B^2} \ (T - T_c) \ , \    
        {\rm and} \ \ \sigma = - {C^2 \over B^3} \ E \ . \]   
   
\subsubsection{First-order transition in the absence of external field}   
   
	For the $\sigma = 0$ case minimization of the Gibbs free energy    
	(\ref{firstfield}) with respect to $\zeta$ implies the equation  
\begin{equation}   
	\frac{\partial \Delta \tilde{\cal G}}{\partial \eta}   
	 	= \tau \zeta - \zeta^2 + \zeta^3 = 0 \ .    
	\label{third-order}   
\end{equation}   
	   
	For $\tau > \tau_0 = {1 \over 4}$ there is the only minimum at    
	$\zeta = 0$ corresponding to a high-temperature undistorted phase.   
	The second minimum at $\zeta = \frac{1}{2}\,\left(1 + \sqrt{1 - 4 
	\tau} \right)$, or
\begin{equation}   
	\eta = - {B \over {2 C}} \left(1 + \left(1 - {4 \alpha C \over  
		B^2}\,(T - T_c^{\prime})\right)^{- \frac{1}{2}} \,\right) 
	\label{FOE}   
\end{equation}   
	appears at $T_0$ and corresponds to a low-temperature distorted phase  
	which initially has higher free energy. The phase energies become    
	equal at $\tau_\ast = {2 \over 9}$, though the supercooling of    
	the high-temperature state as well as superheating of the    
	low-temperature one are possible.    
	It means that the first-order phase transition takes place at the    
	temperature $T_\ast$. As well as in the second-order case,  
	high-symmetry phase becomes unstable at $\tau =0$. 
   
	At the temperature of the first-order transition $T_\ast$ the    
	order parameter jumps from the $\eta = 0$ to the    
\begin{equation}   
	\eta = - {2 \over 3}\,{B \over C},    
\end{equation}   
	overcoming the activation energy barrier    
\begin{equation}   
	\Delta {\cal G}_b = {1 \over 324}\,{B^4 \over C^3}.    
\end{equation}   
	The entropy now has a finite change at transition temperature    
\begin{equation}   
	\Delta {\cal S} = - {2 \over 9}\,{{\alpha B^2} \over {C^2}},    
\end{equation}   
	that correspond to latent heat of the phase transition   
\begin{equation}   
	\Delta {\cal Q} = {2 \over 9}\,{{\alpha B^2} \over {C^2}} T_\ast.   
\end{equation}   
	 
	In order the weakly-first-order transition to be properly identified  
	and clearly separated from the background of critical fluctuations 
	around $T_c$ (\ref{fluct}), the transition jump of the order  
	parameter should be greater than its mean fluctuation. In other word, 
	the energy scale of the problem should be larger than the thermal  
	fluctuation energy $k_B T$. It implies the condition for the value of  
	the third-order coefficient in the Ginzburg-Landau expansion 
\begin{equation}   
	B \ \ \ge \ \  \left( C^3 k_B T_c \right)^{1 \over 4} .   
	\label{Bcond} 
\end{equation}   
	If this condition is not satisfied, the phase transition is "too  
	weak" to be first-order and will be seen in the experiments as a  
	continuous one.

\subsubsection{Effect of the external field on the weakly-discontinuous    
	phase transition}   
\label{fieldfirst}   
   
        Substituting $\zeta = \tilde{\zeta} + {1 \over 3}$ into the     
        Ginzburg-Landau expansion (\ref{firstfield}), we get the     
        third-order term excluded\cite{Fradkin}:    
\begin{equation}    
        \Delta \tilde{\cal G} =     
                {\tilde{\tau} \over 2} \ \tilde{\zeta}^2 +     
                {\tilde{\zeta^4} \over 4} - \tilde{\sigma} \tilde{\zeta} +    
                \Delta \tilde{\cal G}_0\ ,     
        \label{transgl}    
\end{equation}    
        where    
\begin{eqnarray}    
        \tilde{\tau} = \tau - {1 \over 3} \ ; \ \     
	\tilde{\sigma} = \sigma - {\tau \over 3} + {2 \over 27} \nonumber \\  
	{\rm and} \ \ \ \Delta \tilde{\cal G}_0\ = {{\tau} \over 18} -    
	{{\sigma} \over 3} - {1 \over 108}.  
	\label{firstfieldvarbls}  
\end{eqnarray}    
   
        This is equivalent to the free energy expansion (\ref{GLF}) for the  
	second-order phase transition under the external field, the only  
	difference consisting of the term $\Delta \tilde{\cal G}_0$ that is  
	independent on $\tilde{\zeta}$. It appears because the free energy is  
	counted with respect to the $(\zeta = 0)$ or $(\tilde{\zeta} = -  
	{1 \over 3})$ state, that implies $\Delta \tilde{\cal G}_0\ = \Delta  
	\tilde{\cal G}(\tilde{\zeta} = 0) \ne 0$.     
            
        The condition (\ref{minimum}) leads to cubic equation    
	(\ref{field}) with the effective temperature $\tilde{\tau}$ and    
	field $\tilde{\sigma}$ instead of the real ones. The sign of    
	discriminant     
\begin{equation}    
        Q = \left( {\tilde{\tau} \over 3}\right)^3 +     
                \left( {\tilde{\sigma} \over 2}\right)^2     
                \propto  4 \sigma + 27 \sigma^2 -     
                        18 \sigma \tau - {\tau^2} + 4 \tau^3     
\end{equation}    
        of this equation indicates, whether it has one root or three ones    
	in real numbers. The latter case corresponds to the appearance    
	of different minima on $\Delta \tilde{\cal G} (\tilde{\zeta})$,     
        second minima of the free energy appearing when $Q(\tau, \sigma)    
	< 0$.     
    
        Hence, (\ref{transgl}) can be considered as the Ginzburg-Landau     
        expansion for the phase transition between the states, related    
	with different minima of the Gibbs free energy. The minima have    
	non-zero values of order parameter, because the symmetry is broken    
	already by the applied field for any temperature. As there is no    
	symmetry breaking, it is not a true phase transition, described by     
        the Landau theory, however, the undistorted phase with $\zeta = 0$    
	can be treated as an analog of ideal high-symmetry      
        "praphase"\cite{Levanyuk} that would allow the symmetry reduction to  
	both of the phases which provide minima of the free energy. It is   
        interesting to note that the first-order phase transition in absence   
        of external field appears to be equivalent to the second-order one   
        under the action of 'effective' external field, the only feature of  
        this situation is zero value of $\eta$ for one of two minima of   
        $\Delta \tilde{\cal G} (\tilde{\zeta})$.    
    
\begin{figure} 
\epsffile{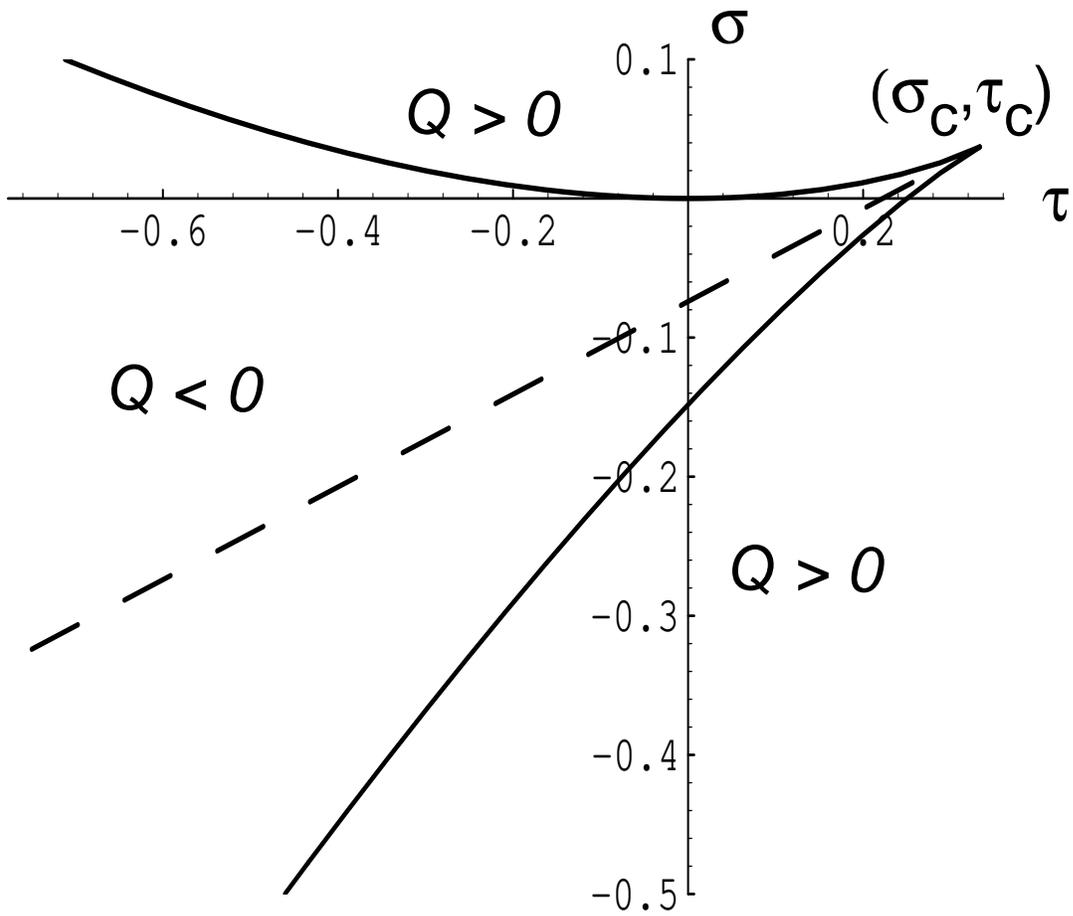} 
\caption{The region of the phase coexistence. The dashed line corresponds to   
        points of the first-order phase transition. It terminates in the  
        critical point $(\tau_c = {1/3}, \sigma_c = {1/27})$.}  
        \label{diagram}  
\end{figure}

        The phase diagram in $(\tau, \sigma)$ plane is shown at the     
        Fig.\ref{diagram}. Additional minimum of the free energy appears    
	for $\sigma_1 \le \sigma \le \sigma_2$ with    
\begin{equation}    
        \sigma_{1,2} = - {2 \over 27} \left(1 \pm (1 - 3\,\tau)^{3\over 2}     
        \right) + {\tau \over 3},     
\end{equation}      
        that leads to the hysteresis with respect to the external     
        field $\Delta \sigma = {4 \over 27} \left( 1 - 3\, \tau    
	\right)^{3\over 2}$.    
            
        According to an analogy with the second-order phase transition    
	described by (\ref{GLF}), the different minima of the $\Delta    
	\tilde{\cal G}(\tilde{\zeta})$ have equal energy only at    
	$\tilde{\sigma} = 0$. This is the condition of the first-order phase   
	transition between corresponding phases and it determines the effect 
	of applied field on the transition temperature $\tau_\ast$     
\begin{equation}    
        \tau_\ast(\sigma) = 3\, \sigma + {2 \over 9} \ .    
        \label{tastsigma}    
\end{equation}    
        For $\sigma = 0$ we get naturally $\tau_\ast(0) = {2 \over 9}$.    
	The Eq.(\ref{tastsigma}) corresponds to the straight line on    
	$(\tau, \sigma)$ plane (Fig.\ref{diagram}). For $\tau > {1 \over    
	3}$ on this line the equilibrium phase has $\zeta = {1 \over 3}$.    
	This state is an analog of the undistorted high-symmetry phase of    
	the Landau theory without an external field which is unstable for    
	$\tau < {1 \over 3}$ and becomes a maximum of $\Delta    
	\tilde{\cal G}$, i.e. the energy barrier with a height of    
\begin{displaymath}   
	{\cal E}_b = {9 \over 4} \sigma^2 - {\sigma \over 6} +    
	{1 \over 324},   
\end{displaymath}   
	for the first-order transition between two different minima with    
	$\tilde{\zeta}_{1,2} = \pm \sqrt{- \tilde{\tau}}$,     
        separated by the order parameter discontinuity     
\begin{equation}    
        \Delta \tilde{\zeta} = \Delta \zeta =  
		{2 \over 3} \sqrt{1 - 27 \sigma} \ . 
	\label{jumpfield}    
\end{equation}    
    
 	As this line of the first-order transition in the phase diagram 
	at $(\tau, \sigma)$ plane separates states without symmetry-breaking   
	relationship\cite{LandLif}, it terminates in critical point $(\tau_c = 
	{1 \over 3}, \sigma_c = {1 \over 27})$. The    
	discontinuity in order parameter as well as the potential barrier    
	separating different minima of the free energy vanishes when    
	approaching this critical point. There is no transition for    
	$\sigma > \sigma_c$ or $\tau > \tau_c$, that means suppressing the    
	weakly-first-order phase transition under the fields greater than    
	the critical one. In a contrast with the second-order case where    
	arbitrary small external field destroys the phase transition, here    
        we find that the fields lower than $\sigma_c$ preserve the transition. 
	The critical point is an analog of the continuous    
	phase transition from state with $\tilde{\zeta} = 0$ corresponding    
	to the breaking of symmetry with respect to change of the 
	$\tilde{\zeta}$ sign.

\section{Proper ferroelastic phase transition}    
     \label{PFE}    
    A phase transition characterized by the appearance of spontaneous strain  
    at the transition temperature is called ferroelastic\cite{Salje}.  
    When this spontaneous strain describes the     
    symmetry breaking at the transition, and is, thus, an order     
    parameter, the proper ferroelastic transition takes place. For the     
    case of improper ferroelastics the spontaneous strain is a     
    complimentary order parameter, coupled with the primary one.     
    
    The free energy difference between the parent and product phases for     
    the case of proper ferroelastic transition is due to the elastic     
    strain and corresponds to the Ginzburg-Landau expansion of the     
    elastic energy in series of the strain components\cite{Bocc}. The     
    second-order term in the Ginzburg-Landau expansion is a linear     
    combination of the second-order elastic constants that vanishes at     
    the critical temperature. It is the eigenvalue of the     
    lattice stiffness matrix corresponding to the relevant irreducible     
    representation of the symmetry group of the high-temperature  
    phase\cite{LandLif}. The strain tensor components transforming with     
    respect to this representation form the order parameter and the     
    phase transition belongs to so-called "soft-mode" class, because it     
    is accompanied by a noticeable softening of the corresponding     
    acoustic mode of atomic vibrations\cite{Cowley}, visible as a central    
    peak of the inelastic neutron scattering.    
    
\subsection{Spontaneous strain in cubic lattice}   
    In what follows the case of cubic symmetry of a high-temperature     
    phase will be considered, that describes the martensitic     
    transformations in the A15 compound\cite{a15ssp} as well as in some     
    metallic alloys. The spontaneous strain tensor has only     
    diagonal components and the order parameter is composed by their     
    symmetrized linear combinations\cite{Bocc}    
\begin{eqnarray}    
     \eta_1 = {1 \over \sqrt{6}} (- \epsilon_{xx} - \epsilon_{yy}    
               + 2 \epsilon_{zz}) 
	\label{eta1} \\    
     \eta_2 = {1 \over \sqrt{2}} (\epsilon_{xx} - \epsilon_{yy}) \ .    
\end{eqnarray}    
	The $\eta_1$ corresponds to the extension along $z$ axis without    
	the volume change and $\eta_2$ describes the strain    
	non-tetragonality in $XY$ plane. The critical acoustic mode is a    
	transverse phonons distributing in $\langle {\rm 110} \rangle$    
	directions.   
   
    	The elastic free energy expansion can be written in    
	the Ginzburg-Landau form\cite{AndBlaunt}    
\begin{equation}    
     \Delta {\cal G} = {A \over 2}\,(\eta_1^2 + \eta_2^2) +    
          {B \over 3} \eta_1 (\eta_1^2 - 3 \eta_1 \eta_2^2 )     
       + {C \over 4} (\eta_1^2 + \eta_2^2)^2.    
     \label{AndBlnt}    
\end{equation}   
	 with the following combination of    
	the elastic constants as the coefficients\cite{Liak}   
\begin{eqnarray}   
	{A \over 2} = {\alpha \over 2}(T - T_c) = {1 \over 2}(C_{11} -    
	C_{12}) \\    
	B = {1 \over {6 \sqrt{6}}} ( C_{111} - 3 C_{112} + 2 C_{123}) \\   
	C = {1 \over 48} ( C_{1111} + 6 C_{1112} - 3 C_{1122} -    
			8 C_{1123}) \ .   
\end{eqnarray}   
   
	Substituting \[ \eta_1 = \eta \sin \theta \ \ {\rm and}    
	 \ \ \ \eta_2 = \eta \cos \theta \ , \] we get the Ginzburg-Landau    
	expansion in the form    
\begin{equation}    
     \Delta {\cal G} = {{\alpha \over 2}(T - T_c)}\,\eta^2 -    
	{B \over 3}\,\eta^3 \sin (3 \theta) + {C \over 4}\, \eta^4.    
\end{equation}   
	The minimization with respect to $\theta$    
	\[\frac{\partial \Delta {\cal G}}{\partial \theta} = - B \eta^3    
	\cos (3 \theta) = 0 \ \ \ {\rm and} \ \ \ \    
	\frac{\partial^2 \Delta {\cal G}}{\partial^2 \theta} > 0 \]    
	implies $\sin (3 \theta) = \pm 1$ depending on the sign of $B$. In    
	the case of $B < 0$ we get three solutions    
\begin{equation}   
 	(\eta, 0) \ , \ \ (- {1 \over 2}\,\eta , {\sqrt{3} \over 2}\,\eta )  
	 \ \ {\rm and} \ \     
	(- {1 \over 2}\,\eta , - {\sqrt{3} \over 2}\,\eta ) \ ,   
	\label{sol3}   
\end{equation}     
	corresponding to the low-symmetry tetragonal phases obtained by the   
	extension of parent cubic lattice along three coordinate axis.    
	The free energy dependence on $\eta$ for these minima is the    
	single-component Ginzburg-Landau expansion (\ref{TOT}) for the    
	weakly-discontinuous phase transition. As the solutions are related    
	through the symmetry transformation from the cubic point symmetry    
	group and, thus, are completely equivalent, we can consider further    
	only one of them, e.g. $(\eta, 0)$, without loss of generality. Hence, 
	we can use the Landau theory for the case of single-component order   
	parameter.   
    
    	The cubic symmetry allows the third-order term in the    
	Ginzburg-Landau expansion to appear, hence, the Landau theory says   
 	that the phase transition should be of the first order. Indeed,    
	for the case of Ni-Al and some other systems partial mode    
	softening takes place and the finite strain appears at the    
	transition, though the shear modulus decreases considerably near    
	the phase transition temperature. Some other so-called    
	"pre-transformation" phenomena were found in a number of    
	alloys\cite{Finl83,Tan,Muto,Barsch}. This case respects to    
	the weakly first-order transition mentioned in Section \ref{WFOT}.     
   
	However, this third-degree term appears to be very small in In-Tl,    
	V$_3$Si and some other alloys\cite{Brass} where the critical mode    
	becomes almost complete soften, i.e. shear modulus $(C_{11} -    
	C_{12})$ vanishes as temperature goes to $T_c$. Central peak of    
	the inelastic neutron scattering as well as other critical   
	phenomena appear and the order parameter undergoes very small    
	change at the critical temperature $T_c$\cite{Sound}. So, the    
	transition is very weakly discontinuous, being sometimes of the    
	second order within the experimental accuracy. The considerable    
	enhancement in the acoustic wave magnitude occurs in the vicinity    
	of the critical temperature $T_c$. Hence, this case corresponds to    
	the the second-order phase transformation considered in Section    
	\ref{second-order}.   
   
\subsection{Effect of external stress on the transition}   
   
	An applied pressure gives rise the 'external' stress tensor    
	$\hat{E}$ corresponding to linear term $ - \hat{\epsilon}\,\hat{E}$    
	in the free energy expression\cite{LanLifEl}. The symmetry-breaking   
	strain components are sensitive only to diagonal components of   
	$\hat{E}$, hence, relevant external stress is superposition of   
	hydrostatic pressure $P = {\rm Tr} (\hat{E})/{\sqrt{3}}$ with ones   
	applied (uniaxially) along $z$ axis    
\begin{equation}   
	E_1 = ( - E_{xx} - E_{yy} + 2 E_{zz})/{\sqrt{6}}    
	\label{uniaxpres} 
\end{equation}   
	and within $XY$ plane \[ E_2 = {1 \over \sqrt{2}}\,(E_{xx} - E_{yy}).  
	\] Hence, $E_1$ and $E_2$ are external fields, conjugated to the  
	primary order parameter components $\eta_1$ and $\eta_2$,  
	respectively, whereas hydrostatic pressure will be shown below  
	affect the ferroelastic phase transition through   
	the volume change $\eta_0$ related with the order parameter by the   
	coupling term proportional to $\eta_0 (\eta_1^2 + \eta_2^2)$.   
   
	Applied non-hydrostatic pressure $E_{1,2}$ breaks the symmetry of    
	undistorted phase, moving out corresponding minimum of $\Delta   
	{\cal G} (\eta_1, \eta_2)$ from the origin and lifts the degeneracy  
	between three low-symmetry phases (\ref{sol3}). The pressure which   
	is co-aligned with the spontaneous strain corresponding to one of   
	these solutions, e.g. $(\eta_1, 0)$, preserves the tetragonal  
	symmetry of distorted phase and the pressure value $E_1$ is an  
	external field conjugated to the value of symmetrized strain  
	$\eta_1$ as a single-component order parameter, considered in  
	Section \ref{fieldfirst}. Otherwise the stable low-temperature state  
	of the system has rhombohedral lattice with three different lattice   
	parameters that is characterized by the pressure-dependent $\eta_1$   
	and $\eta_2$. In what follows the effect of uniaxial pressure $E_1$   
	conjugated to $\eta_1$, which preserves the tetragonal symmetry of   
	low-temperature phase is analyzed.  
  
	In the agreement with the analysis of Section \ref{ExtField} the  
	uniaxial pressure was found\cite{Patel} to suppress the ferroelastic  
	phase transition from cubic to tetragonal lattice in the V$_3$Si  
	compound, where third-order term is very small and the transition is  
	of the second-order. For Ni-Al alloy, where the first-order phase  
	transition takes place, the uniaxial pressure appears\cite{Shapiro93}  
	to shift the transition temperature linearly in complete agreement 
	with the Eq.(\ref{tastsigma}). It should be noted that though this 
	alloy exhibit martensitic transition where spontaneous homogeneous  
	strain is accompanied by so-called shuffle\cite{GodKru-89} related 
	with $q \ne 0$ critical mode, the central peak of the inelastic 
	neutron scattering as well as noticeable softening of $C_{11} - 
	C_{12}$ elastic constant appear well above the transition temperature. 
	Hence, this case can also be analysed in the frame of the Landau 
	theory of ferroelastic phase transition.

\section{Volume change in the elastic energy expansion}   
	\label{VLC}   
   
	If the phase transition is sensitive to applied external hydrostatic  
	pressure then there is a difference in the volume of elementary cell  
	of the parent and product phases. The volume change, indeed, takes  
	place in some cases and it was shown\cite{Good} that virtual  
	volumetric strain could reduce the potential barrier for the system  
	to overcome in the phase transition development. In order to analyze 
	the volume change one needs to include corresponding terms into the   
 	Ginzburg-Landau expansion of the Gibbs free energy. Then the   
	equilibrium state of the system should provide the minimum of the    
	free energy with respect to both shear strain and volumetric one.   
   
\subsection{Linear energy of the thermal expansion}   
	\label{VOLIN}   
   
	In the linear elasticity theory\cite{LanLifEl} the trace of    
	strain tensor is a    
	measure of the volume change, the corresponding symmetrized    
	linear combination of the strain components having a form   
\begin{equation}   
	\eta_0 = {{\rm Tr} (\hat{\epsilon}) \over \sqrt{3}}  =    
	{1 \over \sqrt{3}} ( \epsilon_{xx} + \epsilon_{yy} + \epsilon_{zz}). 
\end{equation}   
	Due to thermal strain as well as external pressure, the expansion    
	of the elastic energy in series of   
	$\eta_0$ should start with the linear terms\cite{LanLifEl}:   
\begin{equation}   
	\Delta {\cal G}_0(\eta_0) = - \kappa_0 A_0 (T - T_R)\eta_0 +    
	{A_0 \over 2}\eta_0^2 + P \eta_0,   
	\label{volume}   
\end{equation}   
	where $T_R$ is some reference temperature for volume change, 
	$\kappa_0$ and $A_0$ are the volume thermal expansion coefficient and 
	bulk modulus, respectively and $P$ is external pressure applied. The 
	bulk modulus $A_0 = {1 \over \sqrt{3}}(C_{11} + 2 C_{12})$ is always   
	positive.   
   
	The minimization of $\Delta {\cal G}_0(\eta_0)$ with respect to  
	$\eta_0$ implies equilibrium values of $\eta_0 (T)$ and  
	$\Delta {\cal G}_0(T,P)$   
\begin{equation}   
	\eta_0 = \kappa_0 (T - T_0) - {P \over A_0}    
	\label{etaT}   
\end{equation}   
\begin{equation}   
	\Delta {\cal G}_0 = - {A_0 \over 2} \alpha_0^2 (T - T_0)^2   
		- {P^2 \over {2 A_0}} + \kappa_0 P (T - T_0)   
	\label{vol}   
\end{equation}   
   
	It is easy to see that usual thermodynamical expressions for   
	isothermal compressibility $\beta_T$ and thermal expansion  
	coefficient    
\begin{equation}   
	\beta_T = - \frac{\partial^2 \Delta {\cal G}_0}{\partial P^2} =    
		- \frac{\partial \eta_0}{\partial P} = {1 \over A_0}   
	\label{comprsbl}  
\end{equation}   
\begin{equation}   
	\kappa_0 = \frac{\partial^2 \Delta {\cal G}_0}{\partial P     
		\partial T} = \frac{\partial \eta_0}{\partial T}  
	\label{expcoef}  
\end{equation}   
	are satisfied for such an expression of the Gibbs free energy if one   
	bears in mind that $P$ is opposite of the pressure {\em inside} the   
	system, which is a thermodynamical variable (we also put the volume  
	of whole system equal to 1 for convenience).   
   
\subsection{Effect of the volume change on ferroelastic phase transition}   
	   
	The lowest-order term of coupling between the shear strains 
	$\eta_{1,2}$ and volume change $\eta_0$ for the case of a cubic  
	symmetry of the high-temperature phase is\cite{Liak}    
\begin{equation}   
	\Delta {\cal G}_{int} = D \eta_0 (\eta_1^2 + \eta_2^2),   
\end{equation}   
	where $D = {1 \over {2 \sqrt{3}}}(C_{111} - C_{123})$. Substituting    
	this term into the expansion of elastic energy in a power series of  
	the volume change, we get the expression (\ref{etaT}) for the 
	equilibrium value of $\eta_0$ in the following form   
\begin{equation}   
	\eta_0 = \kappa_0 (T - T_0) - {P \over A_0} -   
			{D \over A_0}(\eta_1^2 + \eta_2^2) \ .    
	\label{etaNT}   
\end{equation}  
	  
	The coupling term does not induce any anisotropy in the  
	$(\eta_1, \eta_2)$ plane, hence, three equivalent minima of the Gibbs 
	free energy at low temperature have a form (\ref{sol3}). We consider 
	one particular solution of the form $(\eta_1, 0)$. The uniaxial  
	pressure $E_1$ applied along $Z$ axis does not break the tetragonal  
	symmetry of corresponding low-temperature phase. Thus, it is an  
	external field conjugated to the $\eta_1$ order parameter and we can  
	use the results of Section \ref{fieldfirst} to study its influence on  
	the phase transition. The volume change $\eta_0$ as well as its  
	derivatives -- thermal expansion coefficient $\alpha$  
	and isothermal compressibility $\beta_T$ does not depend directly on  
	$E_1$, but only follows the dependence of symmetrized shear strain  
	$\eta_1$ through Eqs.(\ref{etaNT}).  
 
        The general expression for the free energy has the following form  
\begin{equation}   
	\Delta {\cal G} = \Delta {\cal G}_0 + \Delta {\cal G}_1 +    
			\Delta {\cal G}_{int} 
	\label{gen}   
\end{equation}   
	where $\Delta {\cal G}_1$ is the Ginzburg-Landau expansion (\ref{TOT}) 
	with respect to $\eta_1$ as the single-component order parameter with  
	included effect of applied uniaxial pressure $E_1$(\ref{uniaxpres}) 
\begin{equation}   
	\Delta {\cal G}_1 = {\alpha_1 \over 2}(T - T_c)\eta_1^2 +   
	{B_1 \over 3}\eta_1^3  + {C_1 \over 4}\eta_1^4 - E_1 \eta_1 \ .   
	\label{UNRTOT}   
\end{equation}   
	Substituting (\ref{etaNT}) with $\eta_2 = 0$ into Eq.(\ref{gen}) 
	and taking $T_c$ as a reference temperature $T_R$ for the volume  
	change, we get the following renormed Ginzburg-Landau expansion of  
	$\Delta {\cal G}$ in the power series of $\eta_1$   
\begin{eqnarray}   
	\Delta {\cal G}(T,P,\eta_1)  
	= \Delta {\cal G}_0 + {A^{\prime}(T,P) \over 2} \,\eta_1^2 +  
		\nonumber \\ 
	{B^{\prime} \over 3}\, \eta_1^3 +  
	{C^{\prime} \over 4} \,\eta_1^4 - E_1 \eta_1 ,   
	\label{RTOT}   
\end{eqnarray}   
	with the coefficients 
\begin{eqnarray}   
	A^{\prime}(T,P) = \alpha^{\prime}\,(T - T_c^{\prime}) = \nonumber \\ 
	= \left( (\alpha_1 + 2 \kappa_0 D) (T - T_c) -  
				{{2 P D} \over A_0} \right),  
	\label{derivsAp} 
\end{eqnarray}   
\begin{equation}   
	B^{\prime} = B_1 \hspace{8pt} {\rm and} \hspace{8pt}  
	C^{\prime} = C_1  - {{2 D^2} \over A_0} \ . 
\end{equation}   
 	Here $\Delta {\cal G}_0\,(T,P)$ is the energy of a high-symmetry 
	phase with $\eta_1 = 0$, given by the (\ref{vol}). The critical 
	temperature	 
\begin{equation}   
	T_c^{\prime} = T_c + {{2 D P} \over {A_0 (\alpha_1 + 2 \kappa_0 D)}}
	\label{TEMP}   
\end{equation}   
	is shifted by the applied hydrostatic pressure   
\begin{equation}   
	{{dT_c} \over {dP}} = {{2 D } \over {A_0 (\alpha_1 + 2 \kappa_0 D)}}. 
	\label{dtdp}   
\end{equation}   
	in agreement with experimental studies, e.g. in In-Tl  
	alloys\cite{Sound86}. New stiffness is given by the formula 
\begin{equation}   
	\alpha^{\prime} = \alpha_1 + 2 \kappa_0 D   \ . 
\end{equation}   
	In order the Eq.(\ref{RTOT}) be considered as an analog of   
	Eq.(\ref{TOT}), the $\alpha^{\prime}$ and $C^{\prime}$ should    
	be positive.    
   
	The inequality $C^{\prime} < C_1$ corresponds to the smoothening of    
	the potential relief when additional degrees of freedom appear,  
	which allows the system to relax easily. It should be noted that the  
	inclusion of the terms corresponding to the possible volume change  
	into the Ginzburg-Landau expansion (\ref{UNRTOT}) does not affect  
	the third-order term $B_1$, hence, the order of phase transition  
	could not be changed by applied hydrostatic pressure. 	  
 
	As the hydrostatic pressure change the transition temperature 
	according to Eq.(\ref{dtdp}), the line (\ref{tastsigma}) of the 
	first-order phase transition becomes a surface in the 3D $(T,P,E_1)$ 
	phase diagram given by a formula
\begin{equation}  
        T_{\ast}^{\prime}(P,E_1) = T_c + {{2 B_1^2} \over {9 \alpha^{\prime}  
	C^{\prime}}} + {2 D \over {\alpha^{\prime} A_0}} \,P - {3 C^{\prime}  
	\over {\alpha^{\prime} B_1}} \,E_1 \ . 
\end{equation} 
	The transition can be induced by variation of hydrostatic pressure 
	under the fixed values of temperature $T$ and uniaxial pressure $E_1$ 
	at the point 
\begin{equation}  
        P_{\ast}(T,E_1) = {{\alpha^{\prime} A_0} \over {2 D}}\,(T - T_c) -   
                {{A_0 B_1^2} \over {9 D C^{\prime}}} +  
		{3 A_0 C^{\prime} \over {2 D B_1}} \,E_1 \ . 
\end{equation}  
        
	The critical point of the end of transition line (\ref{tastsigma}) is 
	now a line in the $(T,P,E_1)$ phase diagram given by the uniaxial 
	pressure $E_c = - B_1^3/(27 {C^{\prime}}^2)$ and critical hydrostatic 
	one that depends linearly on the temperature  
\begin{displaymath}  
        P_c (T) = {{A_0} \over {2 D}} \left(\alpha^{\prime} (T - T_c) -   
                {{B_1^2} \over {3 C^{\prime}}}\right) 
\end{displaymath}  
        and vanishes when $T$ goes to $T_{cp} = T_c +  
	B_1^2/(3 \alpha^{\prime} C^{\prime})$. It means that some values of  
	hydrostatic pressure suppress the transition, caused by the change of 
	uniaxial pressure $E_1$. The closer temperature is to the $T_{cp}$ the 
	smaller hydrostatic pressure needed for the transition to disappear. 
 
\subsection{Transition anomalies of isothermal compressibility, thermal  
		expansion coefficient and specific heat}  
 
	The isothermal compressibility, thermal expansion coefficient and  
	specific heat are expressed by Eqs.(\ref{comprsbl}), (\ref{expcoef})  
	and (\ref{specheat}), respectively, through the second derivatives of  
	$\Delta {\cal G}(T,P,\eta_1)$ with respect to temperature and  
	hydrostatic pressure for an equilibrium value of $\eta_1$ given by  
	the condition (\ref{minimum}) in the form 
\begin{equation}  
	{\cal F}(A^{\prime}, \eta_1) = A^{\prime}\,\eta_1 + B_1\,\eta_1^2 + 
			C^{\prime}\,\eta_1^3 - E_1 = 0 
	\label{inexpl}   
\end{equation}   
	For the first derivative we have the following formula 
\begin{equation}  
	\frac{\partial}{\partial T}\,\left( \Delta {\cal G} 
	- \Delta {\cal G}_0 \right) =  
	\frac{\partial \Delta {\cal G}}{\partial A^{\prime}}\,\frac{\partial  
	A^{\prime}}{\partial T} + \frac{\partial \Delta {\cal G}}{\partial  
	\eta_1}\,\frac{\partial \eta_1}{\partial T}  
\end{equation}   
	Second term in this expression vanishes as equilibrium $\eta_1$ is  
	given by the Eq.(\ref{minimum}) and taking into account  
	Eq.(\ref{derivsAp}) we get the formula 
\begin{equation}  
	\frac{\partial }{\partial T}  
	 \left( \Delta {\cal G}- \Delta {\cal G}_0 \right) = \alpha^{\prime}\, 
			{\eta_1^2 \over 2}\,, 
\end{equation} 
	that leads to the following expression for the transition anomaly of  
	the specific heat 
\begin{equation}  
	\Delta {\cal C}_P = - T \frac{\partial^2}{\partial T^2}  
	\left( \Delta {\cal G}- \Delta {\cal G}_0 \right) 
	= - {\alpha^{\prime}}^2\,T\, 
	\eta_1\,\frac{\partial\eta_1}{\partial A^{\prime}} 
	\label{specheatgen} 
\end{equation} 
	For isothermal compressibility and thermal expansion coefficient we 
	get analogous expressions 
\begin{eqnarray} 
	\Delta \beta_T = - \frac{\partial^2}{\partial P^2}  
	\left( \Delta {\cal G}- \Delta {\cal G}_0 \right) = 
	- {4 D^2 \over A_0^2}\,\eta_1\,\frac{\partial\eta_1} 
		{\partial A^{\prime}}  
	\label{compsbltygen} \\ 
	\Delta \kappa = \frac{\partial^2}{\partial P \partial T}  
	\left( \Delta {\cal G}- \Delta {\cal G}_0 \right)  
		= - {2 D \alpha^{\prime} \over A_0}\, 
	\eta_1\,\frac{\partial\eta_1}{\partial A^{\prime}} \ . 
	\label{expcoefgen} 
\end{eqnarray} 
	It is easy to see that the Keesom-Ehrenfest 
	relationships\cite{LandLif}
\begin{equation}   
	{{dT_c} \over {dP}} = {{\Delta \beta_T} \over {\Delta \alpha}} =  
	{{T \Delta \alpha} \over {\Delta C_P}}  
	\label{keeserenf} 
\end{equation}   
	are satisfied.   
 
	Differentiating both sides of Eq.(\ref{inexpl}) we get 
\begin{displaymath} 
	\frac{\partial \eta_1}{\partial A^{\prime}} =- \frac{\partial {\cal F}}
	{\partial A^{\prime}} \left( \frac{\partial {\cal F}}{\partial \eta_1} 
	\right)^{-1} 
\end{displaymath} 
	and resolving (\ref{inexpl}) with respect to $A^{\prime}$, we can 
	finally obtain the following expression 
\begin{equation}  
	\eta_1\,\frac{\partial\eta_1}{\partial A^{\prime}} =  
	- {\eta_1^3 \over {E_1 + B_1\,\eta_1^2 + 2 C^{\prime}\,\eta_1^3}} \ . 
	\label{etadeta} 
\end{equation} 
    
\subsection{Second-order case}   
   
	Let us consider the case of the martensitic phase transition of the   
	second order for which $B_1 = 0$. In absence of the uniaxial pressure 
	(\ref{etadeta}) does not depend on $\eta_1$ and, hence, neither on  
	temperature nor on hydrostatic pressure:  
	\[ \eta_1\,\frac{\partial\eta_1}{\partial A^{\prime}} =  
	- {1 \over 2 C^{\prime}} \ . \] 
 
	Low-temperature phase appears at $T_c^{\prime}$ with continuous  
	evolution of the order parameter that gives the volume difference  
	between the parent and product phases  
\begin{equation}   
	\Delta \eta_0 = - {\alpha^{\prime} D \over A_0  
		C^{\prime}}\,(T_c^{\prime} - T)  
\end{equation}   
	and the discontinuities of the isothermal compressibility, thermal   
	expansion coefficient and specific heat at phase transition take the   
	following forms   
\begin{equation}   
	\Delta \beta_T = {2 \over {C^{\prime}}} \left( {D \over A_0} \right)^2 
	 	= \frac{2 D^2}{A_0^2 C_1 - 2 A_0 D^2}   
	\label{cmprsblt2} 
\end{equation}   
\begin{equation}   
	\Delta \kappa = {{\alpha^{\prime}} \over {C^{\prime}}} {D \over A_0} 
		= \frac{(\alpha_1 + 2 \kappa_0 D) D}{A_0 C_1 - 2 D^2}   
	\label{SOTAL}   
\end{equation}   
\begin{equation}   
	\Delta C_P = {T_c^{\prime} \over 2}{{(\alpha^{\prime})^2} \over  
	{C^{\prime}}} =  T_c^{\prime}  \frac{A_0 (\alpha_1 + 2 \kappa_0 D)^2} 
	{2 (A_0 C_1 - 2 D^2)}   
	\label{specheat2} 
\end{equation}   
   
	Thus, to find three independent   
	parameters of the model - $\alpha^{\prime}$, $C^{\prime}$ and   
	${D \over A_0}$, we have four measurable values -- $\Delta C_P$,  
	$\Delta \beta_T$ and $\Delta \alpha$ along with  
\begin{equation}   
	{{dT_c} \over {dP}} = {2 \over \alpha^{\prime}} {D \over A_0},   
\end{equation}  
	which are related by Eq.(\ref{keeserenf}). 
   
	As was mention above, for the case of second-order ferroelastic phase  
	transition considerable critical fluctuations take place due to the 
	softening of $C_{11} - C_{12}$ shear modulus. The inhomogeneous  
	fluctuations $\eta_1({\bf x})$ appear to be relevant only in very  
	close vicinity of the critical temperature\cite{Cowley}. From  
	Eq.(\ref{fluct}) that describes the homogeneous order parameter  
	fluctuations taking into account Eq.(\ref{etaNT}) we get the  
	following expressions for the critical fluctuations of volume  
\begin{equation}	   
	\eta_0 \propto - {D \over A_0} \,|T - T_c|^{-1}    
	\label{fluctvol}  
\end{equation}  
 	and thermal expansion coefficient   
\begin{equation}	   
	\kappa = \frac{\partial \Delta \eta_0}{\partial T} \propto    
			{D \over A_0} \,|T - T_c|^{-2}.   
	\label{fluctexpans}  
\end{equation}   
	in the temperature interval around $T_c$, where the fluctuations  
	in $\eta_1$ are important. 
	   
	There are some experimental data available on the anomalies of thermal 
	expansion in the single-crystal specimens near the martensitic phase   
	transition\cite{SmithFinl}, which are in an agreement with the above   
	results. In most of the alloys the thermal expansion coefficient   
	increases near $T_c^{\prime}$, that implies the positiveness of $D$.   
	Outside the temperature region of thermal fluctuations the thermal    
	expansion coefficient does not depend on the temperature. Besides,    
	linear model of Eq.(\ref{volume}) leads to the independence of 
	$\alpha$ on the applied hydrostatic pressure.  
 
	The second-order phase transition disappears under the applied 
	external field $E_1$, hence, $C_P$, $\beta_T$ and $\kappa$ manifest 
	continuous behavior near the temperature $T_c$ under arbitrary small 
	external field. However, from Eq.(\ref{etadeta}) we get in such a case 
\begin{equation}  
	\eta_1\,\frac{\partial\eta_1}{\partial A^{\prime}} =  
	- {\eta_1^3 \over 2 C^{\prime}}  
	\left( {E_1 \over 2C^{\prime}} + \eta_1^3 \right)^{-1} \ .  
\end{equation} 
	For sufficiently small values $E_1$ of external uniaxial pressure the  
	difference in the isothermal compressibility, thermal expansion  
	coefficient and specific heat outside close vicinity of the transition 
	temperature appears to be described by Eqs.(\ref{cmprsblt2}) - 	 
	(\ref{specheat2}). The critical divergencies given by  
	Eqs.(\ref{fluctvol}) and (\ref{fluctexpans}) disappear and the smeared 
	peaks around $T_c^{\prime}$ appear instead.  
 
\subsection{The first-order transition}   
	   
\subsubsection{Absence of external uniaxial pressure} 
	\label{firsabsfield} 
 
	If the third-order coefficient in the Ginzburg-Landau expansion    
	(\ref{UNRTOT}) has a non-zero value, then the Eq.(\ref{RTOT}) in the  
	absence of external uniaxial pressure  
	describes the first-order phase transition at the temperature   
\begin{equation}	   
	T_\ast^{\prime} = T_c^{\prime} + {2 \over 9}{B_1^2 \over    
		{\alpha^{\prime} C^{\prime}}}   
		= T_c + {{2 D P} \over {A_0 \alpha^{\prime}}} +  
		{2 \over 9}{B_1^2 \over {\alpha^{\prime} C^{\prime}}}  
\end{equation}   
	The shift of the transition temperature by the applied hydrostatic  
	pressure has the same form (\ref{TEMP}) as for the second-order  
	transition. The volume difference between the phases appears to  
	depend on the temperature through the temperature dependence of  
	$\eta_1$ given by Eq.(\ref{FOE}) 
\begin{equation}   
	\Delta \eta_0 = - {{D B_1^2} \over {4 A_0 {C^{\prime}}^2}} 
	\left(1 + \left(1 - {4 \alpha^{\prime} C^{\prime} \over B_1^2} 
	\,(T - T_c^{\prime})\right)^{- \frac{1}{2}} \,\right)^2 \ .  
	\label{voldiff} 
\end{equation}   
	This leads to finite volume change at the phase transition temperature 
\begin{equation}   
	\Delta \eta_0(T_\ast^{\prime}) =  
		- {4 \over 9}{B_1^2 \over (C^{\prime})^2}  
\end{equation}   
	that can be observed in diffraction as well as dilatometric studies.   
   
	Eq.(\ref{etadeta}) takes the form 
\begin{eqnarray}  
	\eta_1\,\frac{\partial\eta_1}{\partial A^{\prime}} =  
	- {\eta_1 \over {B_1 + 2 C^{\prime} \eta_1}} = \nonumber \\ 
	= - {1 \over 2 C^{\prime}} \left(1 + \left(1 -  
	{4 \alpha^{\prime} C^{\prime} \over B_1^2}\,(T -  
	 T_c^{\prime})\right)^{- \frac{1}{2}} \right)	   
\end{eqnarray} 
	and along with Eq.(\ref{expcoefgen}) gives the temperature dependence 
	of difference in thermal expansion coefficients between the low- and 
	high-symmetry phases in the form 
\begin{equation}   
	\Delta \kappa = {{\alpha^{\prime}} \over {C^{\prime}}}   
				{D \over A_0}   
	\left(1 + \left(1 - {4 \alpha^{\prime} C^{\prime} \over B_1^2}\,(T -  
	 T_c^{\prime})\right)^{- \frac{1}{2}} \,\right)	\ .   
	\label{expcoeff1}  
\end{equation}   
	The change of thermal expansion coefficient at the transition point 
	$T_{\ast}^{\prime}$ now has a form
\begin{equation}   
	\Delta \kappa(T_\ast^{\prime}) = {4 {\alpha^{\prime}} \over   
				 {C^{\prime}}} {D \over A_0} =    
		4 D \,\frac{\alpha_1 + 2 \kappa_0 D}{A_0 C_1 - 2 D^2} \ .   
	\label{expcoeff1j}  
\end{equation}   
	The value of $\Delta \kappa$ decreases to that given by  
	Eq.(\ref{SOTAL}) as $T$ goes down from $T_\ast^{\prime}$ to  
	$T_c^{\prime}$.  
   
	For the difference in isothermal compressibility between  
	low-temperature phase and high-temperature one we can find similarly 
\begin{equation}   
	\Delta \beta_T = {{2 D^2} \over {C^{\prime} A_0^2}}   
	\left(1 + \left(1 - {4 \alpha^{\prime} C^{\prime} \over B_1^2}\,(T -  
	 T_c^{\prime})\right)^{- \frac{1}{2}} \,\right)	   
	\label{comprsblty1}  
\end{equation}   
	that gives us the phase transition discontinuity in the form    
\begin{equation}   
	\Delta \beta_T(T_{\ast}^{\prime}) = 
		{{8 D^2} \over {C^{\prime} A_0^2}} \ .   
	\label{comprsblty1j}  
\end{equation}   
	It should always be positive according to general thermodynamical    
	arguments\cite{LandLif}.    
   
	The specific heat has the following temperature dependence  
\begin{equation}	   
	\Delta C_P = {T \over 2}{{{\alpha^{\prime}}^2} \over    
		{C^{\prime}}}    
	\left(1 + \left(1 - {4 \alpha^{\prime} C^{\prime} \over B_1^2}\,(T -  
	 T_c^{\prime})\right)^{- \frac{1}{2}} \,\right)	   
	\label{specheat1}  
\end{equation}   
	with the jump at the transition temperature   
\begin{equation}	   
	\Delta C_P(T_{\ast}^{\prime}) = 2 T_{\ast}^{\prime}   
	{{{\alpha^{\prime}}^2} \over {C^{\prime}}}  = 2 T_{\ast}^{\prime}   
	{A_0 (\alpha_1 + 2 \kappa_0 D)^2 \over {A_0 C_1 - 2 D^2}}	   
	\label{specheat1j}  
\end{equation}   
	The phase transition discontinuities of volume and entropy are related 
	through the Clapeyron-Clausius relationship with the slope 
	(\ref{dtdp} of the equilibrium line at phase diagram  
\begin{equation}	   
	{{dT_c} \over {dP}} = \frac{\Delta \eta_0(T_\ast^{\prime})}   
	{\Delta {\cal S}} = {2 D \over \alpha^{\prime} A_0}.   
\end{equation}   
	   
	Both the thermal expansion coefficient and isothermal compressibility  
	of undistorted phase with $\eta = 0$ do not depend on temperature, so, 
	the expressions (\ref{expcoeff1}) and (\ref{comprsblty1}) describe the 
	temperature dependence of these quantities in the low-symmetry phase  
	that can be observed experimentally below the transition temperature 
	$T_{\ast}^{\prime}$. It should be noted, however, that the effects 
	can be seen for $T_\ast^{\prime}$ being sufficiently far from $T_c$,  
	outside the temperature region where thermal fluctuations \ref{fluct}  
	are important, because the fluctuation-induced singularities of the  
	thermal expansion coefficient as well as the other quantities become  
	larger in critical region than the jumps in their equilibrium values.  
	Thus, the third-order coefficient $B_1$ should satisfy condition  
	(\ref{Bcond}). 
	 
	Such a situation occurs in the case of Ni-Al and some other alloys  
	where shear modulus at $T_\ast^{\prime}$ is soften only slightly  
	-- by 10 $\div$ 20\%, and the experimentally measured temperature  
	dependence of shear modulus can be interpreted as pointing even to  
	negative $T_c$\cite{GodKru-89}. However, for In-Tl system where  
	third-order term appears to be very small and the shear modulus almost 
	vanishes at the transition temperature, the volume discontinuity is so 
	small\cite{Sound86} that it is hidden by thermal fluctuations   
	(\ref{fluctvol}). The similar effect occurs with respect to other  
	discontinuities at the transition temperature $T_\ast^{\prime}$, which 
	is very close to $T_c$.   
 
\subsubsection{The effect of external uniaxial pressure on the anomalies 
		around the first-order phase transition}  
 
	For the case of the first-order phase transition the dependence of the 
	discontinuity in the order parameter on the uniaxial pressure $E_1$  
	follows from general expression (\ref{jumpfield})  
\begin{displaymath}    
        \Delta \eta_1 = - {2 B_1 \over 3 C^{\prime}} \left(1 + 
		{27 {C^{\prime}}^2 E_1 \over B_1^3}\right)^{1 \over 2} \ .    
\end{displaymath}    
	that leads to the following volume change at $T_{\ast}$ 
\begin{equation}   
	\Delta \eta_0(T_\ast^{\prime}) =  
		- {D \over A_0} \Delta (\eta_1^2) =  
		- {4 \over 9}{D B_1^2 \over {A_0 {C^{\prime}}^2}} 
	\left(1 + {27 {C^{\prime}}^2 E_1 \over B_1^3}\right)^{1 \over 2}  
\end{equation}   
 
	Both parent and product phases have $\eta_1 \ne 0$ and $\Delta  
	{\cal G} \ne 0$ under applied external uniaxial pressure, thus, the  
	differences in the isothermal compressibility, thermal expansion  
	coefficient and specific heat are proportional to  
\begin{displaymath}    
	\Delta \left( \eta_1\,\frac{\partial\eta_1}{\partial A^{\prime}}  
		\right) =  
	\eta_{1,1}\,\frac{\partial\eta_{1,1}}{\partial A^{\prime}} -  
	\eta_{1,2}\,\frac{\partial\eta_{1,2}}{\partial A^{\prime}} ,  
\end{displaymath}    
	where $\eta_{1,1}(T,P)$ and $\eta_{1,2}(T,P)$ correspond to two  
	different minima of the free energy given by the different solutions 
	of Eq.(\ref{inexpl}). Using the dimensionless variables  
	(\ref{firstfieldvarbls}) we can write Eq.(\ref{etadeta}) in the form 
\begin{displaymath}    
	\eta_1\,\frac{\partial\eta_1}{\partial A^{\prime}} =  
		- {1 \over C^{\prime}}\,{(\tilde{\zeta} + {1 \over 3})^3 \over 
		{\sigma - (\tilde{\zeta} + {1 \over 3})^2 +  
		2 (\tilde{\zeta} + {1 \over 3})^3}} 
\end{displaymath}    
	At the transition point we have $\tilde{\sigma} = 0$ and  
	$$\tilde{\zeta} = \pm \sqrt{- \tilde{\tau}} =  
		{1 \over 3} \sqrt{1 - 27 \sigma}, $$  
	that gives the expression 
\begin{equation} 
  	\eta_1\,\frac{\partial\eta_1}{\partial A^{\prime}} = {(\tilde{\zeta}  
	+ {1 \over 3})^2 \over {2 C^{\prime} \tilde{\zeta}^2}} \ . 
\end{equation}   
	Taking into account that $\tilde{\zeta}^2$ has the same value  
	$- \tilde{\tau}$ at the transition point for both phases, we can  
	finally obtain 
\begin{equation} 
  	\Delta \left( \eta_1\,\frac{\partial\eta_1}{\partial A^{\prime}}  
		\right) =	 
	- {2 \over C^{\prime}} \left( 1 - 27\,\sigma \right)^{-{1 \over 2}}. 
\end{equation}   
 
	From the Eqs.(\ref{specheatgen}) - (\ref{expcoefgen}) we get the  
	expressions for the phase transition discontinuities of the specific  
	heat, isothermal compressibility and thermal expansion coefficient 
	as follows  
\begin{equation}	   
	\Delta C_P(T_{\ast}^{\prime}) = 2 T_{\ast}^{\prime}   
		{{{\alpha^{\prime}}^2} \over {C^{\prime}}} \left(1 + 
		{27 {C^{\prime}}^2 E_1 \over B_1^3}\right)^{-{1 \over 2}} 
\end{equation}   
\begin{equation}   
	\Delta \beta_T(T_{\ast}^{\prime}) = {{8 D^2} \over {C^{\prime} A_0^2}} 
	\left(1 + {27 {C^{\prime}}^2 E_1 \over B_1^3}\right)^{-{1 \over 2}} 
\end{equation}   
\begin{equation}   
	\Delta \alpha(T_\ast^{\prime}) = {4 {\alpha^{\prime}} \over   
				 {C^{\prime}}} {D \over A_0}  
	\left(1 + {27 {C^{\prime}}^2 E_1 \over B_1^3}\right)^{-{1 \over 2}} \ .
\end{equation}   
 	 
	In the limit of small values of external uniaxial pressure we get the 
	Eqs.(\ref{specheat1j}), (\ref{comprsblty1j}) and (\ref{expcoeff1j}),  
	derived from their temperature dependence in the Section  
	\ref{firsabsfield}. When $E_1$ goes to the value of the critical 
	point $E_c$, these discontinuities diverge as 
	$\propto |E_c - E_1|^{-{1 \over 2}}$. 
   
\section{Transformation from FCC into BCC lattice via spontaneous strain}   
	\label{BCCFCC}   
   
\subsection{FCC -- BCC transformation through the Bain strain}   
	\label{Bain}   
	There is the case of martensitic transformation of especial interest, 
	namely FCC -- BCC transformation in Fe and some ferrous alloys.   
	Since, there is no group-subgroup relationship for the symmetry    
	breaking, the Landau theory is, generally speaking, inapplicable to 
	this case. However, there is an orientational relationship between 
	lattices of the parent and product phase, and the transition could be 
	described in terms of spontaneous strain of so-called Bain 
	type\cite{Royt}.   
   
	The Bain strain is the single-axis shear of the same kind as an order 
	parameter of the ferroelastic phase transition from cubic to 
	tetragonal lattice. It is accompanied by the volume change, that is   
	approximately 1.5\% in the case of pure Fe, where transformation from 
	austenite FCC $\gamma$-phase to ferrite BCC $\alpha$-one takes place at
	910$^o$C. If the lattice periods for austenite and martensite 
	(ferrite) are $a_\gamma$ and $a_\alpha$ respectively, then the strain 
	tensor components have the form   
\begin{equation}   
	\epsilon_{xx} = \epsilon_{yy} = \sqrt{2}\,{a_\alpha \over a_\gamma} - 1
		 \hspace{6pt} {\rm and} \hspace{8pt} \ \ \   
		\epsilon_{zz} = {a_\alpha \over a_\gamma} - 1 \ .   
	\label{BainTens}   
\end{equation}   
   
	The fundamental feature of this case as compared with the above 
	considered phase transition from cubic lattice to the tetragonal one,
	is the fixed value of the spontaneous strain needed to get the 
	symmetry properties of the low-temperature phase. In the above 
	considered case for any non-zero value of the order parameter $\eta_1$ 
 	the symmetry of the lattice was tetragonal, whereas in the case of the 
	FCC -- BCC transformation the peculiar value of $\eta_1$ is needed to 
	get the low-temperature BCC lattice. If we separate the shear strain 
	from the volume change by taking the latter equal to zero, then we get 
	single (and very large) value of symmetrized shear strain (\ref{eta1}) 
\begin{equation}   
	\eta_1 = -\,{{\sqrt{6} \left(\sqrt{2} - 1 \right)} \over 
			{2 \sqrt{2} + 1}} \approx -\,0.256   
\end{equation}   
	Hence, this phase transition is completely different from the    
	continuous ones, which the Landau theory describes, where the value   
	of the order parameter changes with temperature in low-symmetry phase 
	according to the minimization of its Gibbs free energy (\ref{minimum}).
	   
	However, coupling with the volume change $\eta_0$ makes it possible    
	to have the $\eta_1$ variation in low-temperature phase without 
	breaking of its symmetry. Indeed, the strain tensor (\ref{BainTens}) 
	implies the following expressions for the symmetrized combinations 
	used above as the order parameter components   
\begin{eqnarray}      
	\eta_0 = {{(2 \sqrt{2} + 1)} \over \sqrt{3}}\,   
			{a_\alpha \over a_\gamma} - \sqrt{3} \\ 
	\label{eta0kappa}  
	\eta_1 = -\,\sqrt{2 \over 3}(\sqrt{2} - 1)\,{a_\alpha \over a_\gamma} 
	\hspace{8pt} ; \hspace{12pt} \eta_2 = 0 \ ,  
	\label{eta12kappa}  
\end{eqnarray}   
	and we get the relationship between shear strain and volume change in 
	the form   
\begin{equation}   
	\eta_0 = -\,{{2 \sqrt{2} + 1} \over {2 - \sqrt{2}}}\,\eta_1 - \sqrt{3}.
	\label{FE}   
\end{equation}   
	Hence, the variation in value of $\eta_1$ preserves the BCC structure 
	of the low-temperature phase, if $\eta_0$ is changed in such a way 
	that this relationship is satisfied. It should be noted that 
	(\ref{FE}) is meaningful only in restricted region of $\eta_0$ and 
	$\eta_1$. For example, $\eta_1 = 0$ implies unreal result $a_\alpha = 
	0$ from (\ref{eta12kappa}). Thus, (\ref{FE}) is justified only in some 
	vicinity of the transition that is characterized by small volume 
	change $\sqrt{3}\,\eta_0$. 

	Having supposed FCC -- BCC transformation to be ferroelastic one, we 
	should get the minimum of elastic energy for the values of $\eta_0$ 
	and $\eta_1$ obeying the condition (\ref{FE}). Let us study what are 
	the coefficient in the expansion which provide such a minimum. Without 
	careful analysis, it should be noted, however, that the linear 
	approximation used above gives another kind of relationship 
	(\ref{etaNT}) between the shear strain and volume change, thus, 
	non-linear approximation for the thermal expansion energy should be 
	used.

\subsection{Non-linear elasticity for the volume change}   
   
	Non-linearity arises naturally when taking into account large value of
	the strain tensor component. The ${\rm Tr}(\hat{\epsilon})$ for Bain  
	strain in pure iron is approximately three times larger than real 
	value of the volume change for this transformation given by direct    
	multiplication of the lattice periods of low-temperature phase   
\begin{equation}   
	{{\delta V} \over V} = (1 + \epsilon_{xx})(1 + \epsilon_{yy})(1 +    
		\epsilon_{zz}) - 1.   
\end{equation}   
	It could be expressed through the symmetrized combinations $\eta_0$, 
	$\eta_1$ and $\eta_2$ as follows   
\begin{eqnarray}   
	{{\delta V} \over V} = \sqrt{3}\,\eta_0 + \eta_0^2 +    
  			{\eta_0^3 \over {3 \sqrt{3}}} -    
  		{\eta_1^2 + \eta_2^2 \over 2} -  \nonumber \\  
  		- {\eta_0\,(\eta_1^2 + \eta_2^2) \over {2\,\sqrt{3}}} +    
  		{\eta_1\,(\eta_1^2 - 3\,\eta_2^2) \over 3\,\sqrt{6}} \ .   
\end{eqnarray}    

	For the FCC -- BCC transition we have $\eta_2 = 0$ and proper account 
	for the volume change should, thus, involve the terms of higher order 
	in $\eta_0$ and $\eta_1$. Terms of the first and second order in the 
	volume change $\eta_0$ within non-linear approximation have no longer 
	simple relation with the thermal expansion and isothermal 
	compressibility that was obtained in Section \ref{VOLIN}. Similarly, 
	other terms in both $\Delta {\cal G}_1$ and $\Delta {\cal G}_{int}$    
	should be changed. General non-linear elastic energy expansion near 
	the elastic instability with respect to $\eta_1$ now has the form   
\begin{eqnarray}   
	\Delta {\cal G} = L_0\,{\eta_0} +    
	{A_0  \over 2}\,\eta_0^2 + {B_0  \over 3}\,\eta_0^3 +    
			D\,{\eta_0}\,\eta_1^2  \nonumber \\    
	 + {A_1  \over 2}\,\eta_1^2 + {B_1  \over 3}\,\eta_1^3 +    
			{C_1  \over 4}\,\eta_1^4 \ ,   
	\label{NonLinGLE}   
\end{eqnarray}   
	where the coefficients $B_0$ and $C_1$ in highest-order terms should 
	be positive and we again consider the particular expression with 
	$\eta_2 = 0$. 

	The minimization with respect to $\eta_0$ implies  
\begin{equation}   
	{\partial \Delta {\cal G} \over \partial \eta_0} =    
	L_0 + A_0\,\eta_0   
		+ B_0\,{\eta_0}^2 +  D\,{\eta_1}^2  = 0   
\end{equation}   
	that leads to the following relationship between $\eta_0$ and 
	$\eta_1$ in distorted phase with $\eta_1 \neq  0$ :   
\begin{equation}   
 	\left(\eta_0 + { A_0 \over {2 B_0}} \right)^2 + {L_0 
		\over B_0} - {{A_0^2} \over 4 B_0^2}  = 
		- {D \over B_0}\,\eta_1^2 \ . 
	\label{eta0eta1}   
\end{equation}   
	For the high-symmetry phase we have 
\begin{equation}  
 	\eta_0 = -\,{A_0 \over 2 B_0}\,\left( 1 \mp \left( 1 - 
		{4 B_0 L_0 \over A_0^2}\right)^{1 \over 2} \right)
\end{equation}   
  	As there is no co-existing high-symmetry equilibrium states with 
	different values of $\eta_0$, the condition
\begin{equation}   
	A_0^2 - 4 L_0 B_0 = 0   
	\label{cond}   
\end{equation}   
	should be satisfied. Substituting this expression into 
	Eq.(\ref{eta0eta1}), we get    
\begin{equation}   
	\eta_1^2 = - {B_0 \over D}\,\left( \eta_0 + { A_0 \over {2 B_0}} 
		\right)^2   
\end{equation}   
	In order the right-hand side of this expression to be positive    
	condition $D < 0$ must be satisfied, because of positiveness of $B_0$.
  
	Finally, we can get the following expression for the minimum of the 
	free energy (\ref{NonLinGLE})   
\begin{equation}   
	\eta_0 = -\,\sqrt{- {D \over {B_0}}}\,\eta_1 - {A_0 \over {2 B_0}}
\end{equation}   
	and Eq.(\ref{FE}) along with (\ref{cond}) lead to the following    
	relations between the coefficients in the elastic energy expansion    
	(\ref{NonLinGLE})   
\begin{equation}   
	L_0 = 3 B_0   
\end{equation}   
\begin{equation}   
	A_0 = 2 \sqrt{3}\,B_0   
\end{equation}   
\begin{equation}   
	D = - \frac{9 + 4 \sqrt{2}}{3 - 2 \sqrt{2}}\,B_0 \ .	   
\end{equation} 	
	These relations could be, generally speaking, satisfied only in    
	isolated points on phase diagram and the phenomenological approach 
	used in the present study is unable to find their origin. It    
	could be done only in some microscopic theory beyond the    
	scope of the paper. However, as far as these relations are satisfied, 
	we can try to find their consequences for the elastic properties of 
	the system under phase transition.

	Substituting these expressions into the free energy (\ref{NonLinGLE}) 
	and excluding the volume change through Eq.(\ref{FE}) we get renormed 
	expansion of the elastic energy with respect to symmetrized strain 
	$\eta_1$
\begin{eqnarray}      
	\Delta {\cal G} = -\,\sqrt{3}\,B_0  + 
	\left( A_1 + \sqrt{3}\,B_0\,{18 + 8 \sqrt{2} \over 3 - 2 \sqrt{2}}
	\right)\,{\eta_1^2 \over 2} \nonumber \\ + \left( B_1 + B_0\,
	{815 + 580 \sqrt{2} \over 116 - 41 \sqrt{2}}\right)\,{\eta_1^3 \over 3}
 	+ {C_1 \over 4}\,\eta_1^4 \ ,
	\label{martfe}
\end{eqnarray}   
	which can be considered as a Ginzburg-Landau expansion for 
	ferroelastic phase transition. The first term does not depend on 
	$\eta_1$, the critical temperature $T_c$ is 
	defined by the condition
\begin{displaymath}
	A_1 + \sqrt{3}\,B_0\,{18 + 8 \sqrt{2} \over 3 - 2 \sqrt{2}} = 0 \ .  
\end{displaymath}
	and the temperature $T_{\ast}$ of the first-order transition with 
	finite jump in $\eta_1$ and $\eta_0$ is given by an equation 
\begin{displaymath}
	9\,A^3 - 3\,A\,B^2 + C_1\,B^2 = 0
\end{displaymath}
	where $A$ and $B$ are the expressions in brackets of the  first- and 
	second-degree terms in Eq.(\ref{martfe}). In order some transition 
	line to exist on phase diagram, the $B_0$ coefficient should be 
	temperature- and pressure-dependent.

\section{Conclusions}

	We have analyzed the volume change effect on ferroelastic 
	(martensitic) phase transitions and considered the case of cubic 
	lattice of high-symmetry phase as an example. The minimization of 
	elastic energy with respect to hydrostatic strain as a secondary order 
	parameter is shown to renorm the second- and fourth-order coefficients 
	of the Ginzburg-Landau expansion of elastic free energy in powers of 
	symmetrized shear strain. The coupling between shear strain and volume 
	change appears to shift the transition temperature under applied
	external hydrostatic pressure and lead to the finite volume effect of 
	the weakly discontinuous ferroelastic phase transition.

	The isothermal compressibility as well as thermal expansion coefficient
	is shown to diverge near the critical temperature of the second-order 
	ferroelastic phase transition due to the homogeneous fluctuations of
	the order parameter. The difference between their values in parent and 
	product phases outside the fluctuation region appears to be 
	proportional to coupling coefficient. For the case of first-order 
	transition isothermal compressibility and thermal expansion coefficient
	depend on the temperature in the low-symmetry phase according to the 
	square root law. 

	The uniaxial pressure conjugated to the symmetrized shear strain is 
	shown to suppress the second-order transition, leading to the change 
	of divergencies for smeared peaks in the temperature dependencies of 
	isothermal compressibility and thermal expansion coefficient around 
	critical temperature. We have found the first-order transition surface
	at the phase diagram in coordinates of the temperature and hydrostatic 
	as well as uniaxial pressure. This terminates at the line of critical 
	point and the uniaxial pressure of magnitude lower than critical, 
	shifts the transition temperature, but preserves the transition. The 
	critical hydrostatic pressure that suppresses phase transition has 
	linear temperature dependence. The order parameter discontinuity and 
	the volume effect diverge at the critical line as well as the 
	difference in isothermal compressibility and thermal expansion 
	coefficient between the parent and product phases.

	The coupling between the volume change and shear strain is shown to 
	lead to the FCC -- BCC martensitic transformation for some special 
	relations between the coefficients in the free energy expansion. 
	Though some fixed value of the Bain strain is needed to get the 
	low-temperature BCC lattice, the volume change as a secondary order 
	parameter makes it possible to have some temperature variation of the 
	shear strain preserving the BCC lattice and changing its period only. 
	The non-linear expression for the elastic energy of thermal expansion 
	is shown to lead to proper relation between the shear strain and 
	volume change  for the minima of elastic energy.

\acknowledgments   
   
    	The author is grateful to Prof. A.L. Roytburd for useful discussions. 
	This work was supported, in part, by a Soros Foundation Grant awarded 
	by the American Physical Society. The final part of the work has been 
	done at Carleton University under the hospitality of Prof. J. Goldak.

\epsfxsize=6.5in  

\end{document}